\documentstyle[11pt,newpasp,twoside]{article}
\def\edcomment#1{\iffalse\marginpar{\raggedright\sl#1\/}\else\relax\fi}
\reversemarginpar

\begin{document}
\title{Electromagnetic Mechanism for the Origin of Knots in Parsec 
Scale Jets}
 \author{Vladimir I. Pariev}
\affil{Department of Physics and Astronomy, University of Rochester,
Rochester, NY 14627, USA and Lebedev Physical Institute,
Leninsky Prospect 53, Moscow 119991, Russia}
\author{Yakov N. Istomin}
\affil{Lebedev Physical Institute,
Leninsky Prospect 53, Moscow 119991, Russia}
\author{Andrey R. Beresnyak}
\affil{Lebedev Physical Institute,
Leninsky Prospect 53, Moscow 119991, Russia}

\begin{abstract}
We consider the propagation and stability of the helical electromagnetic 
perturbations in a relativistic force-free Pointing flux dominated jet 
(magnetic helix). 
Perturbations are found to dissipate energy in the vicinity of the 
resonant surfaces inside the jet, where the resonance of the perturbation
with the Alfv\'en waves occurs. The energy of the perturbations is
converted into the energy of accelerated particles, which emit 
synchrotron radiation. The intensity and polarization of this radiation 
is modulated with the perturbation. Perturbations with vanishing group 
velocity are the most important and could be associated with the bright knots
in parsec scale jets.
\end{abstract}

The jets from Active Galactic Nuclei are not uniform. They consist 
of a number of bright knots with fainter emission in between.
There is mounting observational evidence 
that jets in many AGNs are dominated by electron-positron pairs. 
Many jets from AGN are highly collimated, stable over distance 
much exceeding their radii, and have 
superluminally moving knots. The hydrodynamic and MHD stability of 
jets was investigated in many studies, which generally find jets 
to be unstable. The jets are surrounded by 
gas. In case of light strongly magnetized force-free jets, the 
presence of surrounding gas changes the stability properties of the jet 
dramatically. The stability of the
electron-positron jets, rotating and moving with relativistic
speed and surrounded by a dense medium was studied by Istomin \&
Pariev~(1994) and Istomin \& Pariev~(1996). Such jets were shown
to be stable with respect to axisymmetric as well as spiral
perturbations. The density of the media
surrounding the jet, $\rho$, should satisfy the condition $\rho
\gg B^2/(4\pi c^2)$ for the inertia of surrounding gas to
stabilize the relativistic jet. On parsec scales, a typical value
of $B \sim 10^{-2}\,\mbox{G}$ gives a proton density $n \gg
0.05\,\mbox{cm}^{-3}$. The characteristic density of gas in
galactic nuclei is $n \sim 10\,\mbox{cm}^{-3}$. Therefore, the
approximation of stationary jet walls used by Istomin~\& Pariev 
(1994, 1996) is well justified. 

Surrounding gas stabilizes light force-free jets. Magnetic helices
can have dominant toroidal magnetic fields and still be stable 
over long distances. Recent observations of the gradients and 
sign of the rotation measure across the jets in several BL~Lac
objects suggest the presence of the intrinsic toroidal magnetic 
field in the jets (Gabuzda \& Murray 2003).
The change of the direction of the magnetic field from toroidal
close to the jet axis to axial on the periphery was inferred from the 
VLBI polarization measurements in blazar 1055+018 (Attridge et al. 1999).
This change is also a characteristic of a spiral magnetic field.

How bright knots can be produced in a force-free magnetic helix~? 
Alfv\'en and fast magnetosonic speeds are equal to speed of light 
in force-free electrodynamics. Therefore, even very relativistic 
flows inside jets are always subsonic. Shocks do not develop in 
the expanding subsonic flow. Thus, knots are not associated with 
shocks. Two possibilities remain then: 1) knots are perturbations 
of magnetic field propagating along the jet; 2) knots are the sites
of particle acceleration, i.e. ``illuminated'' portions of magnetic
helix. It turns out that these two possibilities are not independent:
particle acceleration naturally occurs whenever a generic disturbance
of electromagnetic fields propagates along the jet.
Below we briefly summarize the results of our works 
Beresnyak, Istomin~\& Pariev (2003) and Pariev, Istomin~\& 
Beresnyak (2003), where all relevant details and more extended
discussion can be found.

The energetic
particles accelerated in the central region cannot survive far
along the jet because of the synchrotron losses.
Recent discoveries of X-ray and optical emission from 
a number of jets (e.g., Marshall et al. 2001; Harris et al. 2003)  
make the need for {\it in situ} particle energization
even more essential. It is very tempting to utilize large scale 
electric fields present in a force-free relativistic jet to accelerate 
particles to high energies. However, because of high conductivity 
of plasma the electric field is directed 
perpendicular to the magnetic field and does not readily produce acceleration. 

In a stable jet,
perturbations do not increase with time
($\mbox{Im}\,\omega=0$) or have a small decay rate
($\mbox{Im}\,\omega\approx10^{-2} \mbox{Re}\,\omega$) because of
the resonance with Alfv\'en waves $\omega^\prime=k_\|^\prime c$.
When the specific energy density and pressure of the plasma are
much less than the energy density of the magnetic field, the
Alfv\'en velocity is equal to the speed of light in vacuum $c$
($\omega^\prime$ and $k^\prime$ are the frequency and the wave
vector in the plasma rest frame). The resonance condition is
fulfilled on the specific cylindrical magnetic surface 
in the case of a cylindrical jet. For each
$\omega^\prime$ the position of the resonance surface is different
(it also depends on the discrete azimuthal and radial numbers 
of the eigenmode). 

The source of the disturbances is most likely non-stationary 
processes in the magnetospheres of the black hole and the accretion 
disc in the ``central engine''. Short time variability on scales from 
days to months is actually observed in AGN. The spectrum of this
variability is a broadband as there is no clearly identified 
periodicity. 
As was shown in Istomin~\&
Pariev~(1994, 1996) a standing eigenmodes ($v_{\rm group}(\omega_s)=0$)
generally exist in relativistic jets with toroidal magnetic 
fields, which do not propagate along 
the jet but are only subject to diffuse spreading. While all 
other disturbances, with $\omega\neq \omega_s$, 
propagate away, the disturbances with the frequency close to
$\omega_s$ do not. Their amplitude grows and much exceeds the amplitude
of all other disturbances.
Therefore, the perturbations become close to monochromatic (for a given 
discrete azimuthal and radial numbers) and  
the Alfv\'en resonance surface corresponding 
to the frequency of the standing wave $\omega_s$ is formed.

In the
vicinity of that surface the magnetic and electric fields of the
wave are large. Particles in the jet are accelerated by the
electric field there, drifting away from the region of the strong 
fields and absorbing the energy of the perturbation.
Thus, the stability of the jet is directly related to the
production of energetic particles in the jet: the perturbation 
is damped with a small damping rate. Wave crests of the standing
modes move along the jet and can be identified with the bright 
knots with typical sizes of
the order of the wavelength of the standing wave, which is about
the radius of the jet.

The polarization of
synchrotron emission in knots is very sensitive to the geometry of the 
helical large scale magnetic field. VLBI observations provide evidence
for the polarized emission in knots and in the space between knots.
This is indicative of the existence of a large scale magnetic field all over
the jet rather than concentrated in separate knots. 
The polarization of synchrotron emission of parsec scale jets is being studied
by few groups (Gabuzda 2000; Lepp\"anen at al. 1995; 
Lister 2001). We used observational data on quasars and BL~Lac objects 
to compare with our calculations of polarizations and proper motion of knots 
in our electromagnetic model described above. 
All relativistic effects were taken into account. We found that if the 
emitting particles are distributed uniformly across the jet, observational
data cannot be matched. If the particles are concentrated in the vicinity
of the Alfv\'en resonance surface, the polarization properties are changed
dramatically, and predictions of our model does not contradict to the 
observational data.

\acknowledgements

The authors are grateful to D.C. Gabuzda for helpful discussions. 
This work was done under the partial support of the Russian Foundation
for Fundamental Research (grant number 02-02-16762).
Support from DOE grant DE-FG02-00ER54600 is acknowledged.

\end{document}